\documentstyle[aps,prl,preprint]{revtex}
\begin{document}

\title{Spectral rigidity and eigenfunction correlations at the Anderson transition}
\date{\today}
\author{J.\  T.\  Chalker$^1$, V.\ E.\ Kravtsov$^{2}$,
 and  I.\  V.\  Lerner$^3$}
\address{$^1$Theoretical Physics, University of Oxford, 1 Keble Road,
Oxford OX1 3NP, United Kingdom\\
$^2$International Center for Theoretical Physics, P.O. Box 586,
34100 Trieste, Italy\\ and
Landau Institute for Theoretical Physics, Kosygina str. 2, 117940 Moscow,
Russia\\
$^3$School of Physics and Space Research, University of
Birmingham, Edgbaston, Birmingham~B15~2TT, United Kingdom
}\maketitle \begin{abstract}
The statistics of energy levels for a disordered conductor are
considered in the critical energy window near the mobility edge.
It is shown that, if
critical wave functions are multifractal, the one-dimensional gas 
of levels on the energy axis is {\it compressible}, in the
sense that the variance of the level number in an interval is
$\langle(\delta N)^{2}\rangle \sim \chi \langle N\rangle$ 
for $\langle N \rangle \gg 1$.
The compressibility, $\chi=\eta/2d$, is 
given {\it exactly} in terms of the
multifractal exponent $\eta=d-D_2$ at the mobility edge
in a $d$-dimensional system.
\end{abstract}

An exact determination of energy levels in complex quantum systems is
neither possible, nor desirable. Research in this area 
has focused instead on a statistical description of the spectra of 
such systems, which include: 
compound nuclei, quantum systems whose classical behaviour is chaotic, and
mesoscopic disordered conductors. In many cases, statistical characteristics 
turn out to be universal, i.e.\ independent of most microscopic
details of the system and thus common to different systems which 
belong to the same universality class. In
 the following we shall be concerned with
the specific example of a disordered conductor at an Anderson transition,
for which universal spectral properties are an aspect of critical behaviour
at the transition. For weakly disordered conductors, where  
wavefunctions corresponding to different levels have
spatial overlap, there exists repulsion between the levels close in 
energy. Typically such repelling levels are governed [1--3]
by Wigner--Dyson statistics (WDS) [4] 
which describe correlations between the
eigenvalues in random-matrix ensembles of certain symmetries [5]. 
At the opposite extreme of strong disorder, spatial overlap
between wavefunctions corresponding to different levels is absent, the levels
are uncorrelated and hence governed by Poisson statistics (PS).

Repulsion between neighbouring levels is only one of the
statistical characteristics which are totally
different in WDS and PS. A second is spectral rigidity,
which characterises fluctuations in the
level density on scales large compared to the mean spacing.
 Specifically, consider the variance, 
$\Sigma_{2}(\langle N\rangle)\equiv \bigl<N^2\bigr>- \bigl<N\bigr>^2$,
for level number fluctuations in an energy interval which on average
contains a large number, $\langle N\rangle$, of levels. 
Spectral rigidity is absent for PS:
$\Sigma_{2}(\langle N\rangle)=\langle N\rangle$,
the maximal possible value. On the other hand, 
level fluctuations are strongly suppressed for WDS:
 $\Sigma_{2}( \langle N\rangle)
\sim\ln\langle N\rangle$ [5]; the levels are almost rigid.
It is useful to think of the energy levels as the coordinates of 
a one-dimensional gas on the energy axis.
If this gas has compressibility $\chi$,
one expects $\Sigma_{2}(\langle N\rangle) \sim \chi \langle N\rangle$
for  $\langle N\rangle \gg 1$.
More precisely, we define (for system size $L$)
 \begin{eqnarray}
 \chi
= \lim_{\langle N \rangle \to \infty} \lim_{L \to \infty}
{d\Sigma_{2}\bigl( \langle N\rangle\bigr)\over d\langle N\rangle}\,.
 \end{eqnarray}
Clearly, $\chi=1$ for a system of
 uncorrelated compressible levels (PS), and $\chi = 0$ for a system of
rigid, imcompressible levels (WDS).

The subject we consider here is the existence of an intermediate {\it 
universal} limit for the compressibility (1). Spectral 
statistics in many systems experience a crossover from WDS to PS
as some parameter of the system is changed. A smooth
crossover of this type takes place for spectral statistics of electrons in
 $2d$ or quasi $1d$ disordered conductors with increasing
disorder.   
A new statistical regime [6,7] arises for $d>2$ at the
critical value of disorder
that corresponds to the Anderson metal-insulator transition. 
The value of the spectral compressibility at the transition was
first considered in Ref.\ [8] and has been a subject
of controversy [7--10]. Note, however, that
all the previous discussion has been limited to the framework of
one-parameter scaling.

We will show here that the one-parameter scaling is insufficient
for the description of level statistics in the vicinity of
the Anderson transition.
The main result of the present work is an
exact expression for the spectral compressibility
at a mobility edge:
\begin{eqnarray}
\chi={\eta\over 2d}\,.
\end{eqnarray}
Here $\eta\equiv d-D_2$ is one of the exponents characterising
properties of the critical eigenstates: the set of
 multifractal dimensions, $D_p$, govern the scaling behaviour of the moments
of inverse participation ratio, 
\begin{eqnarray}
\left<\int d^d r |\psi_n(\bbox{r})|^{2p}
\right>\propto L^{-D_p (p-1)}\,,
\end{eqnarray}
 for integer $p$.
It is only the fractal
dimensionality $D_2$ that enters Eq.\ (2), 
because the compressibility
depends only on the two-level correlation function (TLCF), which (as
we show below) can be
expressed solely via second-order wavefunction correlations.

Now we turn to the derivation of the result, Eq.\ (2). First, we 
express $\chi$ via
the form factor of the TLCF. Then we outline a
non-perturbative approach to
calculating the form factor, developed previously in
Refs.\ [11,12], 
which relates the spectral and wavefunction correlations.
Finally,
in the critical regime near the Anderson tranisition we relate the form factor
to multifractal properties of the wavefunctions. 

We define the TLCF as
 \begin{eqnarray}
 R(s)=\langle \rho \rangle^{-2} \Bigl\{\bigl<\rho(\varepsilon +s\Delta)\,\rho(\varepsilon )
\bigr> \Bigr\} -1\,,
 \end{eqnarray}
where $\rho(\varepsilon )=L^{-d}\sum_n \delta(\varepsilon
-E_n)$ is the density of states per unit volume, $L$ is the sample size,
and $E_n$ are eigenvalues of the Hamiltonian for a system of 
free electrons in a random potential.
The mean  density of states, $\left<\rho\right>$, varies only on the
scale of the Fermi energy, $\varepsilon_{\!_F}$.
Thus the mean level spacing
$\Delta=1/(\left<\rho\right>L^d)$  is considered constant.

The number variance in
an energy window of width $E\equiv \langle N\rangle
\Delta \ll\varepsilon_{\!_F}$ is given in terms of the TLCF by $
 \Sigma_{2}\bigl( \langle N\rangle\bigr)
=\int_{-\left< N\right>}^{\left< N\right>}\bigl(\left< N\right>-
|s|\bigr)R(s)\, ds.$
Differentiating with respect to $\langle N\rangle$,
and taking the limit $\langle N \rangle \to \infty$
{\it after} the thermodynamic limit, $L \to \infty$,
one obtains [7] 
\begin{eqnarray}
\chi=\int_{-\infty}^{+\infty}R(s)\,ds \equiv \lim_{t\to 0}K(t)\,,
\end{eqnarray}
where we have introduced the spectral form factor of the TLCF:
\begin{eqnarray}
K(t) = \int_{-\infty}^{\infty}
e^{-i s t/t_{\!_H }}R(s)\, ds\,,
\end{eqnarray}
and $t_{\!_H }\equiv\hbar/\Delta$ is the Heisenberg time.
It is, of course, only in the thermodynamic limit that one
can expect universal behaviour. Taking this limit at fixed
$ \langle N\rangle $ leaves $ \Delta $ and  $t_{\!_H }$, respectively, 
as the sole energy and time scales. Note that, for a system 
at a mobility edge, the number of critical levels (those
for which the localisation length is bigger than the system
size) diverges in the thermodynamic limit, as is required if
there is to be a possibility that $\chi$ takes a non-trivial 
critical value. We comment in passing that
a well-know sum rule, $\int_{-\infty}^{+\infty}R(s)\,ds =0$,
can be derived if one takes limits in the opposite order:
$\langle N \rangle \to \infty$
{\it before} the thermodynamic limit, $L \to \infty$.
The fact that the order of limits does not necessarily commute
has been discussed extensively elsewhere [9,10,13].

To obtain $\chi$, we use a recently developed  [11,12] 
non-perturbative approach to calculating $K(t)$ for times
$t\ll t_{\!_H }$. In this approach, $K(t)$ is given by
\begin{eqnarray}
K(t)=
{1\over2}\;\;{|t|\,p(t)\over {\pi\hbar\rho +
\int_{0^+}^{|t|}\,p(t') d t'}} \,.
\end{eqnarray}
Here $p(t)=\left<|\Psi({\bbox{0}},t)|^2\right>$ is the
ensemble-averaged quantum return probability 
for a wavepacket, $\Psi({\bbox{r}},t) = \sqrt{V_{0}}\sum_n \psi_n^*(\bbox{0})
\psi_n({\bbox{r}}) e^{-iE_n t / \hbar}$, 
that initially occupies a small
volume  $V_0$ near the origin [14]. The initial condition implies that
the summation here is limited to the number of levels ${\cal N}_0\sim L^d/V_0$
with energies lying within the energy band of width $E_0\sim 1/\rho V_0$.
Using spatial and spectral homogeneity, one can show that 
\begin{eqnarray}
p(t)
&=&\left<\sum_{\,l\alt{\cal N}_0} c_{n,n+l}
e^{-i(E_n-E_{n+l})t/\hbar}\right>\,,\qquad
c_{n,m}\equiv \int \!\!d^dr\, |\psi_n(r)|^2|\psi_{m}(r)|^2\,.
\end{eqnarray}
Thus $p(t)$ and ultimately $K(t)$, Eq.\ (7), are related to the
local wavefunction correlations.

Before  applying Eq.\ (7) to find $\chi$, let us outline
its derivation. A central idea is to consider the dependence of 
energy levels on
some external parameter, $\lambda$. 
Then, Eq.\ (7) can be obtained in two ways  [11,12].
The first is phenomenological: we
consider [11]  a random walk in the ensemble of impurity configurations
parametrised by $\lambda$.
Such a Brownian-motion level dynamics in the fictitious time
$\propto \lambda^2$ is similar to that originally
introduced by Dyson [15] in the context of random matrix theory.
These dynamics generate the level correlations given by Eq.\ (7).
Alternatively [12], using the homogeneity of level
correlations at different points of the parametric space, one can prove
for $t \not= 0$ the identity
\begin{eqnarray}
t p(t) = \int_0^{t} dt'  \left<  \sum_{lm} c_{n,n+l}\,
e^{i(E_n-E_{n+l})t'/\hbar} \,e^{-i(E_n-E_{n+m})t/\hbar} \right>\,,
\end{eqnarray}
which relates $p(t)$ to 
higher-order correlations between energy levels  and wavefunctions
(we stress that in constrast to Eq.\ (8), the summation here is
extended over {\it all} eigenvalues of the Hamiltonian).

In the limit  $t \ll t_{\!H}$ it is possible to factorise
the average on the r.h.s.\ of Eq.\ (9). 
Consider first the sum over $l$.
In both the metallic and the critical regimes,  wavefunctions corresponding
to different energies have spatial overlap, and so $c_{n,n+l}$ is a slowly
 decreasing function of $|l|$. Thus the oscillating factor,
$e^{-i(E_n-E_{n+l})t'/\hbar}$, acts as a cut-off for the sum over
$l$: the number of terms
contributing is of order  $t_{\!H}/t' \agt t_{\!H}/t \gg 1$. As a result,
in the limit $t\to 0$,
the entire contribution comes from terms for which
$|l| \gg 1$.
Now suppose that
for one of these terms (i.e.\ fixed $l$, with
$|l| \gg 1$) we carry out the sum on $m$.
The result will depend on the conditional distribution of $E_{n+m}$, given
that there are levels at the energies  $E_n$ and $E_{n+l}$.
Moreover, for $t \not= 0$, non-zero contributions arise only from
energy regions in which $E_{n+m}$ has a non-uniform distribution due to 
correlations with $E_{n}$ and $E_{n+l}$.
The energy scale for such correlations is set only by $\Delta$, and so the
contribution to Eq.(10) comes from the two 
regions: $|E_{n+m}-E_{n}|\le \Delta$ and $|E_{n+m}-E_{n+l}|\le \Delta$.
Both sets of correlations make equal contributions to the r.h.s.\ of
Eq.\ (9), as can be seen 
by changing variable in the integral from $t'$ to $t''\!=\!t\!-\!t'$.
Taking this into account, we factorise the average in Eq.\ (9) as
\begin{eqnarray}
tp(t) = {2}\left<\sum_m e^{-i(E_n-E_{n+m})t/\hbar}\right>
\int_0^t dt' \left<  \sum_l c_{n,n+l}
e^{i(E_n-E_{n+l})t'/\hbar} \right> \,,
\end{eqnarray}
This is exact for $t/t_{\!H} \to 0$, 
provided that $c_{n,n+l}$ does {\it not}  rapidly decrease
with $|l|$ [16].

The first average in Eq.\ (10) reduces to $K(t)$ on using 
Eqs.\ (4) and (6). 
The second average in Eq.\ (10) differs 
from the return probability 
(8) by a remainder arising from the terms with $l\agt {\cal N}_0$. 
When $|E_n-E_{n+l}|\agt E_0$ the two wavefunctions are essentially
uncorrelated, and $c_{n,n+l}=L^{-d}$. The remainder is therefore
$$
\left<\sum_{\,l\agt{\cal N}_0} c_{n,n+l}
e^{-i(E_n-E_{n+l})t/\hbar}\right>=\left<\int_{E>E_0}\rho(E)
e^{-iEt/\hbar}\right>= 2\pi\hbar \rho \delta_0(t)\,,
$$
where $\delta_0(t)$ is a $\delta$-like function which is zero for
$t\agt t_0=\hbar/E_0$. As $t_0/t_{\! H}$
goes to zero in the thermodynamic limit considered, the integral in
Eq.\ (10) is equal to $\pi\hbar\rho +  \int_{0^+}^{t} p(t') \,dt'$.
Thefore, the factorisation (10) of the
expression (9) leads to the relation (7) between
$K(t)$ and $p(t)$ which is exact in the limit $t/t{\!_H}\to 0$. 

Now we apply  Eq.\ (7) to find the compressibility at the mobility
edge. This regime is characterised by anomalous diffusion. This means that
the diffusion coefficient $D$ becomes time (and scale) dependent. In
the absence of multifractality, it has been shown
[17] by extending one-parameter scaling arguments 
to treat time- or frequency-dependent transport that
$D(t)\propto t^{-1+2/d}$. Then 
(restoring dimensional factors) $p(t)\sim (Dt )^{-d/2}\sim \hbar \rho/t $.
The numerator in Eq.\ (7) is therefore constant while the denominator is 
divergent in the universal limit, $L \to \infty$ with $t/t_{\!_H}$ fixed. 
Thus, in this case, $K(t)=0$ for 
$t\ll t_{\!_H}$ and the compressibility (5) would be zero at the 
mobility edge, as in the metal.

However, there is 
no {\it a priori} reason to extend one-parameter scaling in this 
way. In fact, wavefunction correlations 
at the mobility edge are characterised by a set of
multifractal exponents [18,19]. 
The relevant fractal 
dimension for $p(t)$ is $D_2$ [20], Eq.\ (3),
as it is expressed via $|\psi_n|^2\,|\psi_m|^2$, according 
to Eq.\ (8).
Multifractality reveals itself in the time-dependence of $p(t)$. Since
for $l\ll E_{0}/\Delta$ the factor $c_{n,n+l}\sim |E_n - 
E_{n+l}|^{-\eta/d}$, we have [20]: 
$$
p(t)\sim {V_{0}^{-\eta/d}}\left({\hbar\rho/ t}\right)^{1-\eta/d}
\,,\qquad\qquad
\eta\equiv d-D_2\,.
$$
On subsituting this into Eq.\ (7), we note that the first, constant term in
the denominator is negligible in the limit $L\rightarrow\infty$, with
$t/t_{\!H}$ fixed. Thus we obtain $K(t)=2\eta/d$ for $t\ll t_{\!_H}$. 

Note that in this limit $R(s)=0$ for $s\gg1$, in agreement 
with the diagrammatic results of Ref.\ [7]. There are small
$t$-dependent corrections
to the leading behaviour of $K(t)$ that govern the large-$s$ asymptotics 
of $R(s)$. These corrections, which will  be discussed
elsewhere, do not affect the limiting value
$K(0)$: thus our main result, Eq.\ (2), follows immediately.

This result can be checked directly for $d=2$. In this case, there is no 
Anderson transition but a crossover from the weak to strong localisation 
regime. It is known that, in the  weak  localisation regime, 
where the dimensionless conductance,
$g_{0}=2\pi^{2}\rho D\gg 1$,
the high  moments of the inverse participation ratio exhibit
behaviour characterstic of that for critical states in $d>2$. For the 
orthogonal, unitary and symplectic ensembles labelled by $\beta=1,2$ or $4$,
the corresponding set of multifractal dimensions has been recently found 
non-perturbatively [21] to be $D_p=2-p/\beta g_0$,
in agreement with the earlier renormalization group results [18,22].
This leads to $\eta=2/\beta g_0$, and thus the limiting value of 
compressibility (2) is $\chi=1/2\beta g_0$. On the other hand, the same  
value of compressibility  has been obtained  in this regime with the 
help of direct diagrammatic calcultations [9] (the corresponding
Eq.\ (15) in Ref.\ [9] contains an extra factor of 2 due to
explicit inclusion of spin degeneracy). 

Another possibility [23] to check Eq.(2) is provided by a $1d$ system 
with random hopping integrals $t_{ij}$ whose variance decreases as a
power of  the distance $|i-j|$: 
$\langle(t_{ij})^{2}\rangle=A|i-j|^{2\alpha}$. This system 
is critical at $\alpha=1$, and eigenstates
are multifractal. Calculations 
similar to those described above for 2D systems show [23] that
Eq.\ (2) is again exactly satisfied.

Finally, the relation (2) between the compressibility $\chi$ and the 
multifractality exponent 
can be checked with the help of independent numerical calculations
of both these quantities.  Let us discuss available 
 numerical results. For the 3D Anderson model (which belongs
to the orthogonal symmetry class), the multifractality exponent
$D_{2}$ was found [24,25] to be $D_{2}=1.7\pm 0.2$. 
Using Eq.\ (2) we have $\chi=0.22\pm 0.03$ which is
in a good agreement with the numerical result $\chi=0.2\div0.3$ 
[8,26]. The exponent $\eta$ has been studied 
numerically also for the Anderson transitions in 2D systems belonging to 
the unitary (quantum-Hall regime) and the symplectic symmetry classes [25,27].
In both cases $\eta\approx 0.4$ turns out to be small. Thus
we can predict that the level compessibility should be very 
small,  $\chi\approx 0.1$,  in the vicinity of these transitions. This
could explain that in direct numerical simulations in the 
quantum-Hall regime [28] no
linear contribution to $\Sigma_2$ has been found.   

In summary, we have derived an explicit relationship between the 
compressibility of energy levels and the multifractal  
statistics of eigenfunctions in a disordered system near the Anderson 
transition. A non-zero compressibility implies that  the 
one-parameter scaling is inapplicable to level statistics; for a hypothetical
system where the one-parameter scaling description of the conductance 
could be extended to level correlations, no spectral compressibility
would arise at the transition. 
The  relationship (2) holds exactly for 2D
critical states and is in a reasonable agreement with existing numerical
results.

{\bf Acknowledgements} We thank G.\ Montambaux, L.\ Schweitzer and 
I.\ K.\ Zharekeshev for helpful discussions on numerical results. 
Support from EPSRC grants GR/GO 2727 (J.T.C.) 
and GR/J35238 (I.V.L.) is gratefully acknowledged.
Two of us (V.E.K.\ and I.V.L.) gratefully
acknowledge the hospitality of the ITP in Santa Barbara extended to us
at different stages of this work
and partial support by the NSF under grant No.\ PHY94-07194.


\begin{references}
\baselineskip=14pt

\bibitem{1}
L.~P. Gor'kov and G.~M. Eliashberg,  {Zh.\ Eksp.\ Teor.\ Fiz.\ } {\bf 48}, 1407
  (1965) [{Sov.\ Phys.\ JETP} {\bf 21}, 940 (1965)].

\bibitem{2}
K.~B. Efetov, Adv.Phys. {\bf 32}, 53 (1983).

\bibitem{3}
B.~L. Altshuler and B.~I. Shklovskii,  {Zh.\ Eksp.\ Teor.\ Fiz.\ } {\bf 91},
  220 (1986) [{Sov.\ Phys.\ JETP} {\bf 64}, 127 (1986)].

\bibitem{4}
E.~P. Wigner, Proc. Cambridge Philos. Soc. {\bf 47}, 790 (1951);
F.~J. Dyson, J.\ Math.\ Phys.\ {\bf 3}, 140 (1962).

\bibitem{5}
M.~L. Mehta, {\em Random matrices} (Academic Press, Boston, 1991).

\bibitem{6}
B.~I. Shklovskii, B.~Shapiro, B.~R. Sears, P.~Lambrianides, and H.~B. Shore,
  Phys.\ Rev.\ B {\bf 47}, 11487 (1993).

\bibitem{7}
V.~E. Kravtsov, I.~V. Lerner, B.~L. Altshuler, and A.~G. Aronov, Phys.\ Rev.\
  Lett.\ {\bf 72}, 888 (1994);
A.~G. Aronov, V.~E. Kravtsov, and I.~V. Lerner, {\it ibid} {\bf 74},
  1174 (1995).

\bibitem{8}
B.~L. Altshuler, I.~K. Zharekeshev, S.~A. Kotochigova, and B.~I. Shklovskii,
  {Zh.\ Eksp.\ Teor.\ Fiz.\ } {\bf 94}, 343 (1988) [{Sov.\ Phys.\ JETP} {\bf
  67}, 625 (1988)].
 
\bibitem{9}
V.~E. Kravtsov and I.~V. Lerner, Phys.\ Rev.\ Lett.\ {\bf 74}, 2563 (1995).

\bibitem{10}
A.~G. Aronov and A.~D. Mirlin, Phys.\ Rev.\ B {\bf 51}, 6131 (1995).

\bibitem{11}
J.~T. Chalker, I.~V. Lerner, and R.~S. Smith,  Phys.\ Rev.\ Lett., {\bf 77}
554 (1996).

\bibitem{12}
J.~T. Chalker, I.~V. Lerner, and R.~S. Smith,  
J.\ Math.\ Phys., {in press} (1996).
 
\bibitem{13}
V.~E. Kravtsov, in:  Proceedings of the 1996 Moriond Conference
(cond-mat/9603166).


\bibitem{14}
It is convenient to choose $V_0 \sim \ell^d$, where $\ell$ is the elastic mean
  free path.
 
\bibitem{15}
F.~J. Dyson, J.\ Math.\ Phys.\ {\bf 3}, 1191 (1962).
 
\bibitem{16}
The conditions necessary to derive Eq.\ (10) are not satisfied in the
  insulator, where the dominant contiribution to the r.h.s.\ of Eq.\
  (9) comes from the $l=0$ term. Retaining only this term, Eqs.\
  (9) and (8) yield correctly $K(t)=1$ and hence $\chi=1$.
 
\bibitem{17}
B.~Shapiro and E.~Abrahams, Phys.\ Rev.\ B {\bf 24}, 4889 (1981);
Y.~Imry, Y.~Gefen, and D.~J. Bergman, Phys.\ Rev.\ B {\bf 26}, 3436 (1982).

\bibitem{18n}
F.~Wegner, Z.\ Phys.\ B {\bf 36}, 209 (1980).

\bibitem{19}
C.~Castellani and L.~Peliti, J.\ Phys.\ A {\bf 19}, L429 (1986).
 
\bibitem{20}
J.~T. Chalker and G.~J. Daniell, Phys.\ Rev.\ Lett.\ {\bf 61}, 593 (1988);
J.~T. Chalker, Physica {\bf A 167}, 253 (1990);
B.~Huckenstein and L.~Schweitzer, Phys.\ Rev.\ Lett.\ {\bf 72}, 713 (1994).
 
\bibitem{21}
K.~B. Efetov and V.~I. Falko, Europhys.\ Lett.\ {\bf 32}, 627 (1995).
 
\bibitem{22}
B.~L. Altshuler, V.~E. Kravtsov, and I.~V. Lerner,  {Zh.\ Eksp.\ Teor.\ Fiz.\ }
  {\bf 91}, 2276 (1986) [{Sov.\ Phys.\ JETP} {\bf 64}, 1352 (1986)].

\bibitem{23}
A.~D. Mirlin, Y.~V. Fyodorov, F.-M. Dittes, J.~Quezada, and T.~H. Seligman,
cond-mat/9604163, 1996.
 

\bibitem{24}
C.~M. Socoulis, and E.~N. Economou, {Phys.\ Rev.\ Lett.\ } {\bf 52}, 565 
(1984); 
M. Schreiber, {Physica A} {\bf 167}, 188, (1990); 
S.~N. Evangelou, {J.\ Phys.\ A} {\bf 23}, L317 (1990). 

\bibitem{A}
T. Brandes, B. 
Huckestein, and L. Schweitzer, cond-mat/9605062, 1996. 


\bibitem{B}
D. Braun and G. Montambaux, {Phys.\ Rev.\ B} {\bf 52}, 13906 (1995);
I.~Kh. Zharekeshev and B. Kramer, {Jpn.\ J.\ Appl.\ Phys.\ }, {\bf 34}, 
4361 (1995);
 B.~R. Sears and H.~B. Shore, unpublished (1994).

\bibitem{C}
S.~N. Evangelou, {Phys.\ Rev.\ Lett.}  {\bf 75}, 2550 (1995).

\bibitem{D}
M. Feingold, Y. Avishai, and R. Berkovits, {Phys.\ Rev.\ B} {\bf 52}, 
8400 (1995).
\end{references}
\end{document}